# Confinement Specific Design of SOI Rib Waveguides with Submicron Dimensions and Single Mode Operation †

Abdurrahman Javid Shaikh [1,2,*], Abdul Ghani Abro [1], Mirza Muhammad Ali Baig [1], Muhammad Adeel Ahmad Siddiqui [3], and Syed Mohsin Abbas [4]

[1] Department of Electrical Engineering, NED University of Engineering & Technology, Karachi; arjs@neduet.edu.pk; ghani@neduet.edu.pk; mohdali@neduet.edu.pk.
[2] Haptics Human Robotics and Condition Monitoring Lab, NCRA, NEDUET, Karachi; arjs@neduet.edu.pk
[3] Technical Planning Department, K-Electric, Karachi.
[4] Department of Nuclear Engineering, Pakistan Institute of Engineering and Applied Sciences (PIEAS), Islamabad.
* Correspondence: arjs@neduet.edu.pk.
† Presented at 7th International Electrical Engineering Conference, Karachi, and 26th March, 2022.

**Abstract:** Full-vectorial finite difference method with perfectly matched layers boundaries is used to identify the single mode operation region of submicron rib waveguides fabricated using silicon-on-insulator material system. Achieving high mode power confinement factors is emphasized while maintaining the single mode operation. As opposed to the case of large cross-section rib waveguides, theoretical single mode conditions have been demonstrated to hold for sub-micron waveguides with accuracy approaching 100%. Both, the deeply and the shallowly etched rib waveguides have been considered and the single mode condition for entire sub-micrometer range is presented while adhering to design specific mode confinement requirements.

## 1. Introduction

The first step in designing many of the photonic devices incorporating waveguides is to design a waveguiding structure that supports the fundamental mode only, i.e. the waveguide should be a single mode one [1-4]. For the desired performance, in addition to the single mode requirement, most of the devices require the power confined by the waveguide to have a benchmark value [5]. For large dimensions (height and width of the order of 5μm) SOI rib waveguides, the single mode condition (SMC) has been analytically shown to hold [6], as follows:

$$\frac{W}{H} \leq \alpha + \frac{r}{\sqrt{1-r^2}}; \qquad (1)$$

where W, H and r are rib width, height, and slab height to rib height ratio (r=h/H) as shown in Figure 1.

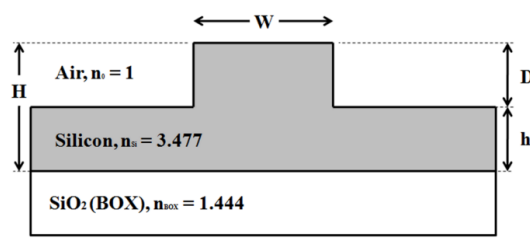

**Figure 1.** A typical rib waveguide structure ($n_0$, $n_{Si}$ and $n_{BOX}$ are real refractive indices of corresponding layers).

Various works [7, 8] supported the same relationship as of (1) with $\alpha$=0.3, with the exception of Pogossian [9] who reported the same value of $\alpha$ to be zero. This change is very significant, suggesting at least 30% variation between the two formulae. Apart from several other studies on the design of large single mode, the work of Lousteau et al. [10] suggested based on rigorous finite-difference (FD) and beam propagation methods (BPM) that no equation or dimensional conditions can be suggested for the design of large single mode rib waveguides. Their results meant that it was impossible to know beforehand, either analytically or intuitively, the dimensional range of a larger rib waveguide where it would support only the fundamental mode, thereby rendering device optimization impossible. However, with the ever-pursued quest of reducing the dimensional features of photonic devices for better utilization of wafer real estate, waveguide devices with the cross-sectional size below a micrometer have been demonstrated. This opens up a new arena of waveguide devices working in subwavelength regime. In this work it has been shown that it is possible to give a couple of conditions which, if fulfilled, guarantees single mode operation of submicron waveguides confining an acceptable amount of mode power.

## 2. Methodology

An industry standard simulation tool based on finite-difference frequency domain method employing perfectly matched layers (PML) was used to calculate the complex propagation constant of the waveguide structures under study. Using advanced scripting environment of the simulation tool, a routine was developed which analyze the optical mode profile supported by the sub-micrometer sized rib structure as shown in Figure 1. The analysis was done at the center of the rib in both vertical and horizontal directions as shown in Figure 2. A single mode waveguide would only support the fundamental mode profile for either a TE mode or a TM mode or both, depending on exact waveguide dimensions. Apart from separating single mode and multimode (MM) dimensions, the routine also stored the fraction of power confined under the rib, and the nature of the mode (quasi-TE or quasi-TM) for each of the single mode waveguides. The dimensions were chosen in this way so as to cover every possible combination in which a practical sub-micrometer waveguide could be fabricated. The simulation window size was optimized so that the results converge as well as the memory and computational requirements remain under acceptable limits. For all waveguide structures, the wavelength of operation was 1550 nm.

## 3. Results and Discussion

The results obtained in this work are compared with other works [11, 12] to validate the procedure conducted to obtain the results. The general agreement in the results validate the procedure used in this work [5].

The main result of analysis has been presented in Figure 3. All the submicron dimensions supporting only the single mode have been plotted with a unique set of axes ($\zeta$H vs 'W'). Other works [6, 7, 9] have been using a standard set of axes (W/H vs h/H) to express their results which is thought to give a boundary between single-mode and multimode of waveguide operation. This approach, however, did not include any account for optical mode confinement factor and, moreover, was later shown to be invalid for large rib waveguides [10]. In this work, the parameter $\zeta$ is defined as;

$$\zeta = \frac{r}{\sqrt{1-r^2}} , \qquad (2)$$

where 'r' is normalized rib height (h/H). Shaded portions in Figure 3 (marked 'A', 'B', and 'C') represent the single mode regions with a particular level of minimum optical mode confinement factor ($\Gamma_{min}$).

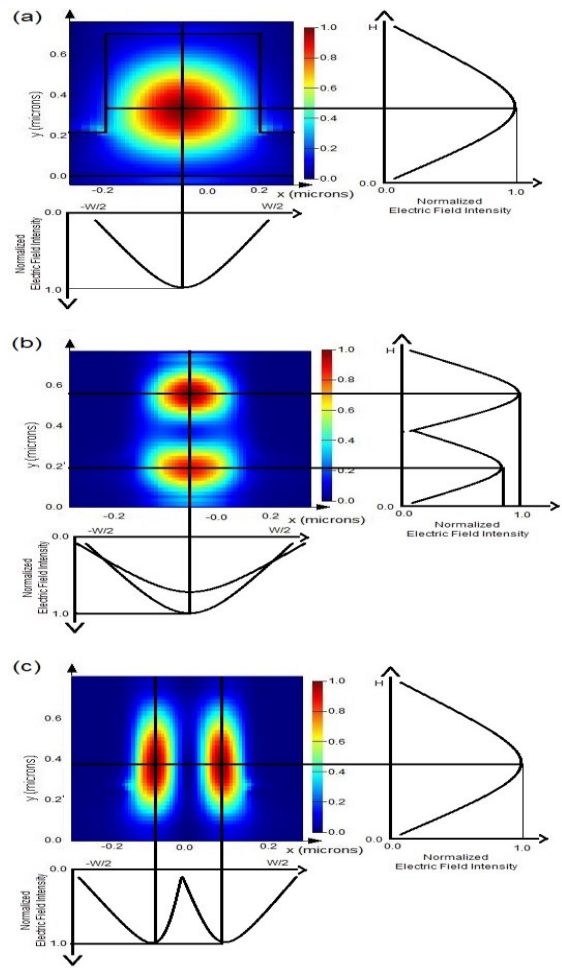

**Figure 2.** Mode field profile under the rib and its respective intensity variation with respect to the vertical (right hand side to the mode profile) and horizontal (below the mode profile) directions through the mode peaks (a) Fundamental mode (b) First-order vertical mode (c) First-order horizontal mode. A single mode waveguide would only support fundamental mode profile.

Reprinted by permission from American Scientific Publishers: J. Nanoelectron. Optoelectron. [5], Copyright 2017.

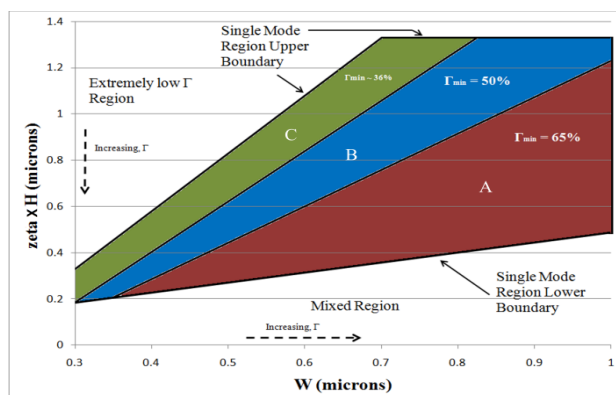

**Figure 3.** Plot of ζH vs waveguide width where the shaded areas contain 100% single modes, with respective confinement factors. The broken arrows show the directions in which Γ increases.

Reprinted by permission from American Scientific Publishers: J. Nanoelectron. Optoelectron. [5], Copyright 2017.

Any point in a particular region of Figure 3 represents a design which confines at least the percentage (given by $\Gamma_{min}$) of total power propagated in the mode volume. It is clear from the figure that as rib width increases and the slab height (h) decreases (manifests by the decreasing values of $\zeta H$ product) the confinement factor would increase. This is true as increasing etch depths (i.e. decreasing 'h') "squeezes" the lower portion of fundamental mode into the rib region by providing an interface to reflect from near lower edges of the rib. For this to happen the rib width should be appropriately large, otherwise, the mode would start to leak out of the silicon region. For a fix slab height, the confinement factor also increases significantly with increasing widths and heights of the rib. Figure 4(a) shows transverse-electric (TE) mode confinement factor of the most commonly used rib height of 220 nm while Figure 4(b) shows the same for 400 nm against the slab height. Here rib width has been used a parameter. It is evident from the figures that slab height affects mode confinement more drastically when the rib height is smaller hence stringent process control is required for fabrication of smaller devices. Secondly, the maximum amount of mode power confined by a particular structure is inversely proportional to the rib height. At the same time, to increase device integration density and operation speeds of active photonic devices, smaller dimensions are required [5]. Figure 4 shows that achieving this is possible should a waveguide is designed with careful selection of rib width and slab height.

Due to its widespread use, the TE-confinement factor relationship with slab height is provided for H=220 nm for different rib width in Table 1.

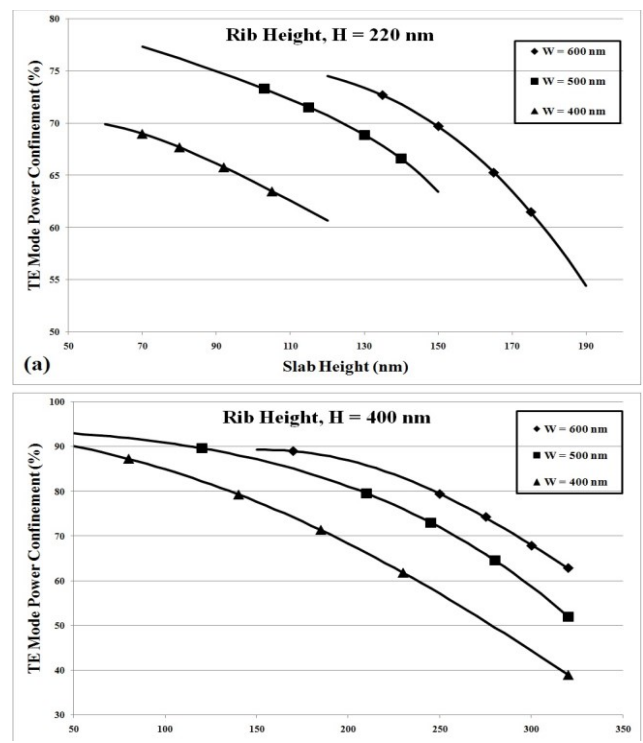

**Figure 4.** TE-mode confinement factor versus slab height for various rib widths; for, (a) H = 220 nm, and (b) H = 400 nm.

## 4. Conclusion

Analysis and design of submicron rib waveguides keeping an account for the mode confinement factor is done in this work. Confinement trends have also been provided in mathematical form for the most commonly used rib height enabling confinement estimates and optimization of devices requiring single mode operation with high confinement factors.

**Table 1.** Relationships of TE confinement factor with respect to waveguide dimensions for the curves shown in Figure 4.

| | H=220 nm |
|---|---|
| **W = 600 nm** | $\Gamma_{TE} = \dfrac{67.8 - 0.257h}{1 - 5.521 \times 10^{-3} + 11.03 \times 10^{-6} h^2}$ |
| **W = 500 nm** | $\Gamma_{TE} = \dfrac{85.04 - 0.472h}{1 - 4.258 \times 10^{-3} - 6.08 \times 10^{-6} h^2}$ |
| **W = 400 nm** | $\Gamma_{TE} = \dfrac{55.42 + 1.09h}{1 + 6.988 \times 10^{-3} + 84.62 \times 10^{-6} h^2}$ |
| | H=400 nm |
| **W = 600 nm** | $\Gamma_{TE} = \dfrac{74.82 - 0.1h}{1 - 3.269 \times 10^{-3} + 7.10 \times 10^{-6} h^2}$ |
| **W = 500 nm** | $\Gamma_{TE} = \dfrac{93.76 - 0.22h}{1 - 2.287 \times 10^{-3} + 1.74 \times 10^{-6} h^2}$ |
| **W = 400 nm** | $\Gamma_{TE} = \dfrac{93.53 - 0.202h}{1 - 1.621 \times 10^{-3} + 2.58 \times 10^{-6} h^2}$ |